\documentclass[reprint,aps,amsmath,amssymb,amsfonts]{revtex4-2}

\usepackage{graphicx}
\usepackage{dcolumn}
\usepackage{bm}
\usepackage{xcolor}

\begin{document}

\newcommand{\TMCT}{T_\text{\tiny MCT}}
\newcommand{\tw}{t_\text{w}}
\newcommand{\TSEB}{T_\text{\tiny SEB}}

\title{Strong ergodicity breaking in dynamical mean-field equations for mixed $p$--spin glasses}

\author{Vincenzo Citro}
\affiliation{DIIN, Universit\'a di Salerno, Via Giovanni Paolo II 132, 84084, Fisciano, Italy and CNR - Nanotec, unit\`a di Roma, P.le Aldo Moro 5, 00185 Rome, Italy}

\author{Federico Ricci-Tersenghi}
\affiliation{Dipartimento di Fisica, Sapienza Università di Roma, Istituto Nazionale di Fisica Nucleare, Sezione di Roma I, and CNR - Nanotec, unit\`a di Roma, P.le Aldo Moro 5, 00185 Rome, Italy}

\date{\today}

\begin{abstract}
The analytical solution to the out-of-equilibrium dynamics of mean-field spin glasses has profoundly shaped our understanding of glassy dynamics, which take place in many diverse physical systems.
In particular, the idea that during the aging dynamics, the evolution becomes slower and slower, but keeps wandering in an unbounded space (a manifold of marginal states), thus forgetting any previously found configuration, has been one of the key hypotheses to achieve an analytical solution. 
This hypothesis, called weak ergodicity breaking, has recently been questioned by numerical simulations and attempts to solve the dynamical mean-field equations (DMFE).
In this work, we introduce a new integration scheme for solving DMFE that allows us to reach very large integration times, $t=O(10^6)$, in the solution of the spherical 3+4--spin model, quenched from close to the mode coupling temperature down to zero temperature.
Thanks to this new solution, we can provide solid evidence for \emph{strong ergodicity breaking} in the out-of-equilibrium dynamics on mixed $p$--spin glass models.
Our solution to the DMFE shows that the out-of-equilibrium dynamics undergo aging, but in a restricted space: the initial condition is never forgotten, and the dynamics take place closer and closer to configurations reached at later times.
During this new restricted aging dynamics, the fluctuation-dissipation relation is richer than expected.
\end{abstract}

\maketitle

Aging dynamics characterize the evolution of a large class of complex systems in their out-of-equilibrium regime. Understanding these processes is crucial for spin glasses, providing valuable insights into the nature of the low temperature phase, the glassy states, and many other applications \cite{bouchaud1998out,berthier2011theoretical,parisi2020theory,cugliandolo2023recent,charbonneau2023spin}.

Mean-field spin-glass models have played a crucial role in our understanding of aging dynamics and enhanced our comprehension of critical phenomena such as history dependency, strong sensitivity to external parameter variations, and relaxation processes spanning all time scales. Among mean-field models, spherical spin glasses are analytically solvable even in some complicated cases \cite{crisanti1992spherical,crisanti1993spherical}. These models describe systems of $N$ spins interacting through random infinite-range $p$--spin interactions
\begin{equation}\label{Hmpsin}
{\cal H} = - \sum^{\infty}_{p=2} \sum_{j_1<...<j_p} \alpha_p \,J_{j_1...j_p} \prod_{k=1}^p \sigma_{j_k}\;,
\end{equation}
where spins $\sigma_i$, $i=1,..,N$ are real variables subjected to the global constraint $\sum_{i=1}^N s_i^2=N$, and the couplings $J_{j_1...j_p}$ are independent Gaussian variables with variance $\overline{J_{j_1...j_p}^2}=p!/2N^{p-1}$. 
Here, we adopt the convention to indicate disorder averaging by an overline, while thermal averaging is indicated by angular brackets. The disorder in the couplings $J_{j_1...j_p}$ is quenched, meaning that all statistical equilibrium and out-of-equilibrium averages are computed with a fixed realization of these couplings. The coefficients $\alpha_p$ define the structure of the Hamiltonian. The model is called pure if there is only one non-zero coefficient; otherwise, it is named mixed.

Cugliandolo and Kurchan \cite{cugliandolo1993analytical} (CK) made a ground-breaking contribution to the study of the out-of-equilibrium dynamics of pure spherical $p$--spin models. They introduced an analytical solution of the dynamical mean-field equations (DMFE), which provided new insights into the evolution of disordered systems towards their lowest energy states. The core of this theoretical framework revolved around the concept of weak ergodicity breaking (WEB), which refers to the system's ability to disregard any configuration it reaches within finite time intervals \cite{bouchaud1992weak}. In particular, the pure spherical $p$--spin model forgets its initial state after a sufficiently long time. This infinite memoryless aging process quickly became a fundamental concept in the physical understanding of spin glasses. Moreover, the characteristics of this aging phase were found to be consistent with experimental observations.

Recently, we have seen the accumulation of results suggesting that the WEB may not be valid beyond the pure spherical $p$--spin model. In Ref.~\cite{bernaschi2020strong}, the aging dynamics of a mean-field spin glass model with Ising variables have been studied. Very large sizes and times were chosen to ensure the dynamics stay in the out-of-equilibrium regime, while allowing for safe extrapolations in the large time limit. The main result of Ref.~\cite{bernaschi2020strong} is the measurement of non-zero correlation between the configuration at time $\tw$ and the asymptotic state ($t\to\infty$) of the relaxing system. Such a correlation is found to grow with $\tw$, thus suggesting the aging dynamics get more and more confined in a subregion of the configuration space, while aging. This scenario is called \emph{strong ergodicity breaking} (SEB).
The above result may be questioned because it is based solely on numerical simulations, and it is important to complement it with more analytical arguments.

A class of models whose out-of-equilibrium dynamics can be analytically computed via the solution to the integral-differential equations is precisely the one defined in Eq.~(\ref{Hmpsin}).
An accurate numerical solution of their DMFE up to very large times would assess the general validity of the WEB hypothesis.
In particular, Ref.~\cite{folena2020rethinking} considered the mixed $p$--spin model, which differs from the pure $p$--spin model by the fact that the Hamiltonian is not a homogeneous function of its variables. The authors successfully integrated system dynamics, using a standard finite-difference discretization of DMFE on a uniform grid, for times up to the order of $10^3$. Over this time range, they observed remarkably interesting physical properties for the mixed $p$--spin model, showing significant differences from the pure $p$--spin case. However, to refute the WEB assumptions made by CK would require extending the integration to even longer times. The classical integration procedure was not able to shed light on the asymptotic behaviour of the system, because integrating the {\color{blue} DMFE} over a time range spanning up to $10^6$ would lead to RAM usage of the order of $10^9$ GB. This limitation represents the real bottleneck that has hindered the investigation of the asymptotic regime of mixed models for decades.

Kim and Latz~\cite{kim2001dynamics} proposed a strategy to extend the numerical integration to longer times, achieving results up to $10^6$ by reformulating the DMFE in terms of the integrated response function and adopting discretization on a variable grid. However, this approach remains stable only for quenching to high temperatures and for pure models.
Mannelli et al.~\cite{mannelli2020marvels} adopted an alternative approach dubbed the dynamic grid algorithm, i.e., the discretization step is periodically doubled and the old solution is projected onto the new coarse grid. However, Folena \cite{folena2020thesis} demonstrated a clear relationship between the time dependence of the error, occurring after each contraction, and the stability of the correlation. Once the relative error exceeds the threshold of $10\%$, the correlation consistently diverges from the reference solution. This divergence occurs relatively early, and even with large grids, integration times beyond $10^3$ cannot be achieved. 

In this context, it is evident that a stable, efficient, and low-memory numerical algorithm would shed light on the unexplored asymptotic dynamics of mixed $p$--spin models. Driven by this goal, we have developed a new numerical scheme based on a structured non-uniform mesh that takes advantage of grid refinement in regions where the numerical solution has large gradients \cite{citro2024thesis}. The classical formulation of DMFE in the thermodynamic limit involves the two-time correlation $C(t,t') = \frac{1}{N} \sum_{i=1}^N \overline{\langle \sigma_i(t)\sigma_i(t')\rangle}$ and the linear response function $R(t,t') = \frac{1}{N} \sum_{i=1}^N \frac{\partial \overline{\langle \sigma_i(t)\rangle}}{\partial h_i(t')}$, which fully describe the mean-field disorder-averaged dynamics.
Here, $h_i$ is an external magnetic field added to the DMFE to compute the linear response. Following Kim and Latz \cite{kim2001dynamics}, we rewrite the DMFE in terms of the integrated response function: $F(t,t')=-\int_{t'}^t ds R(t,s)$. This choice allows the Dirac delta function in the response evolution equation to be replaced by a constant, making the solution smoother.
The resulting DMFE describing the evolution at temperature $T_f$ of a system starting at $t=0$ from an equilibrated configuration at inverse temperature $\beta=1/T$ can be rewritten as follows
\begin{widetext}
\begin{eqnarray*}
\partial_t C(t,t') &=& -\mu(t)C(t,t') + \int_0^t ds f''(C(t,s))\partial_s F(t,s) C(s,t')+\int_0^{t'}ds f'(C(t,s))\partial_s F(t',s)+\beta f'(C(t,0))C(t',0)\;,\\
\partial_t F(t,t') &=& -1 -\mu(t)F(t,t') + \int_{t'}^t ds f''(C(t,s))\partial_s F(t,s)F(s,t')\;,\\
\mu(t) &=& T_f+\int_0^t ds f''(C(t,s))\partial_s F(t,s)C(t,s)+\int_0^t ds f'(C(t,s))\partial_s F(t,s)+\beta f'(C(t,0))C(t,0)\;,
\end{eqnarray*}
\end{widetext}
where $f(q)=\sum_p \alpha_p^2 {\color{blue}q^p}/2=\left.\overline{\mathcal{H}(\sigma)\mathcal{H}(\tau)}\right|_{\sigma\cdot\tau=qN}$ is the covariance of the Hamiltonian at a given overlap. 
The energy density is given by $E(t)=-\int_0^t f'(C(t,s))R(t,s)ds-\beta f(C(t,0))$. These equations are valid for any pair of times $t'$ and $t$, which stay finite while taking the $N\to\infty$ limit. The correlation function must satisfy $C(t,t)=1$ (because of the spherical constraint) and $\lim_{t' \to t^\pm} \partial_t C(t,t')=\pm 1$, while for the integrated response function must hold that $\lim_{t'\to t^-} F(t,t')= 0$.

\begin{figure}[t]
\includegraphics[width=0.8\columnwidth]{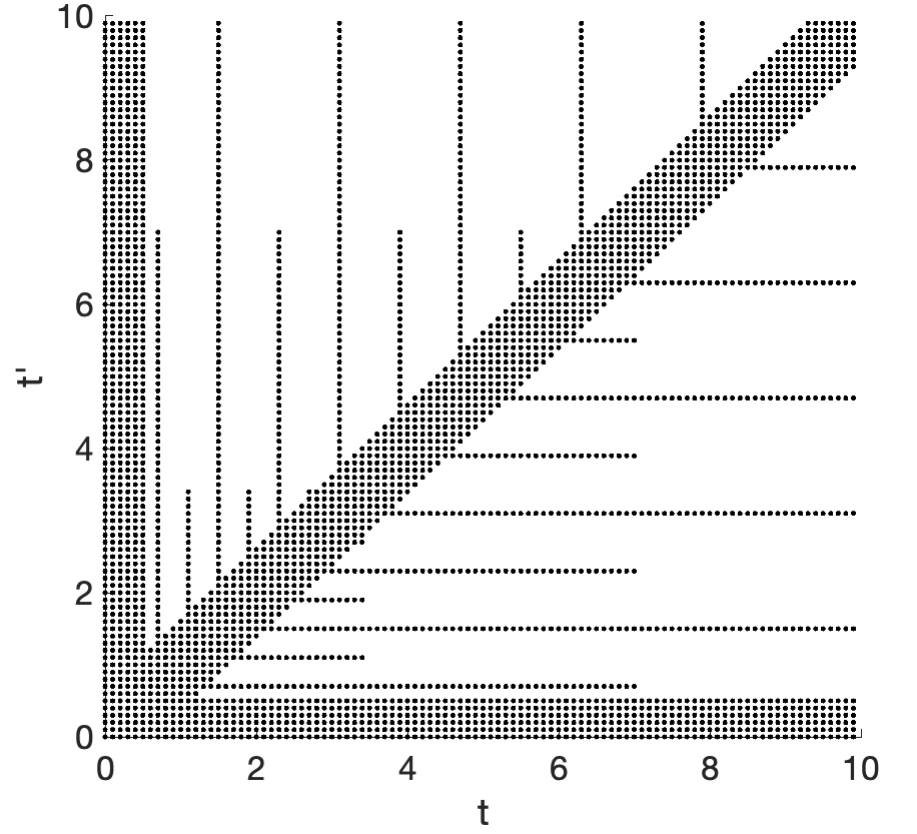}
\caption{The grid in the $(t,t')$ plane used to integrate the DMFE in the present work is a subset of a regular grid of step $\Delta t$. Actually, only the half with $t'\le t$ is evaluated. We choose the integration grid to be dense in the bands close to $t'=0$ and $t'=t$, as there the gradients are larger. The grid becomes sparser and sparser with increasing time, away from the dense bands.}
\label{fig:grid}
\end{figure}

In Fig.~\ref{fig:grid} we show the new grid that we have used for integrating the DMFE in $C(t,t')$ and $F(t,t')$ (only one half of the grid is actually computed). The integration grid is a subset of a regular grid of step $\Delta t$, which becomes sparser at larger times.
We have chosen to keep the integration grid dense along the two bands around $t'=0$ and $t'=t$, because in these regions the gradients are larger, and increasing the integration step would produce systematic errors and integration instabilities.
At the same time, we know that away from these two bands, the gradients become small when times increase, and keeping a dense grid would just represent a waste of memory and computational time.
For this reason, we opt to keep only a subset of points in the internal part of the integration grid, with a spacing that grows proportional to the times. In this way, the number of internal integration points scales roughly linearly with the maximum integration time (instead of quadratically as in a dense grid), and the bottleneck represented by the memory storage is largely solved.
The integration results still depend on a few parameters fixed by the user: the width of the bands and the elementary integration step, $\Delta t$. The dependence on these two parameters has been studied in detail and will be discussed in a future publication fully dedicated to explaining in detail the new scheme of integration for the DMFE.
The results presented in this work have been verified to be stable concerning the specific choice of parameters.

\begin{figure}[t]
\includegraphics[width=\columnwidth]{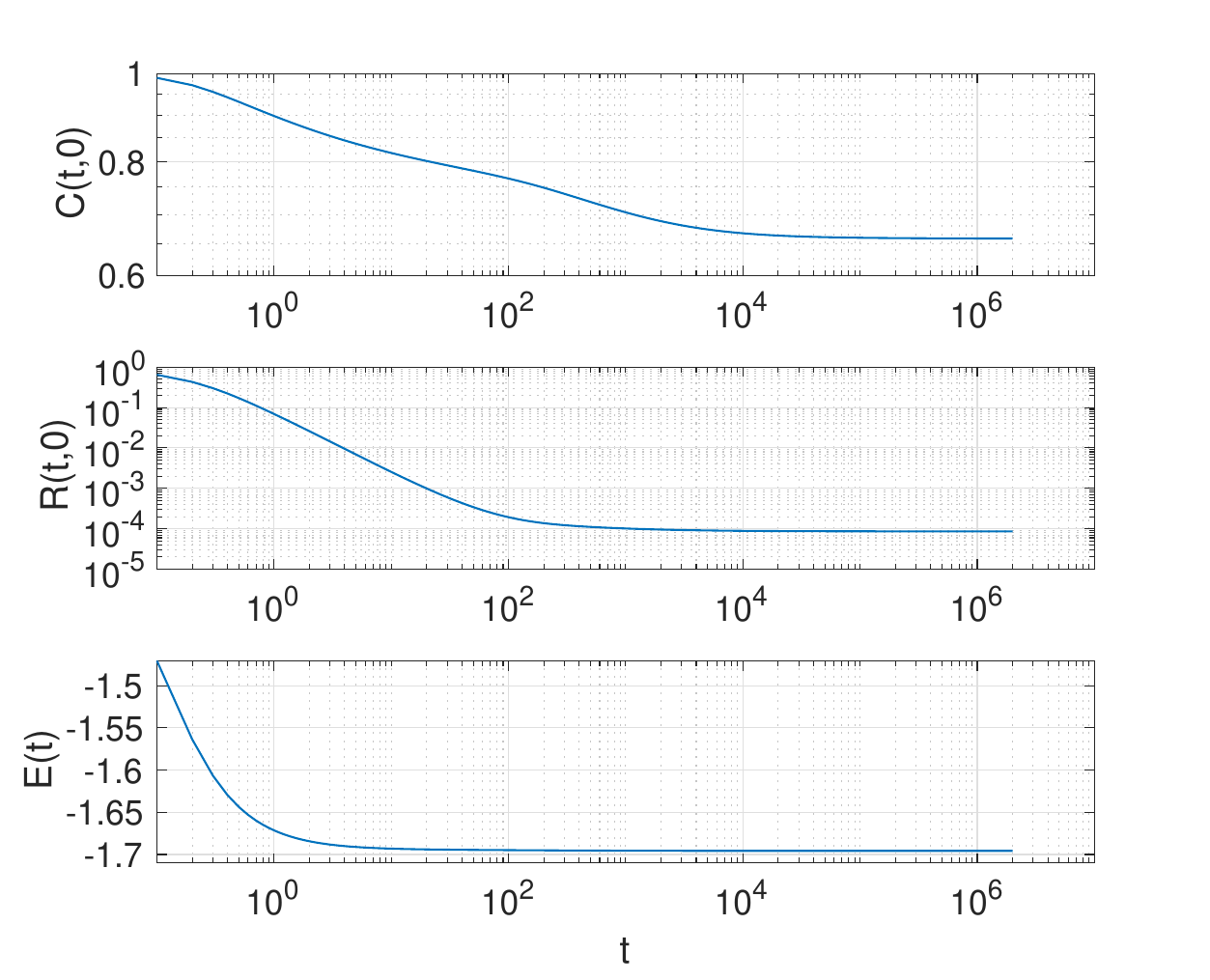}
\caption{Decay of the correlation with the initial condition $C(t,0)$, the evolution of the response $R(t,0)$ and energy relaxation $E(t)$ as a function of time. All the plots are obtained for the $(3+4)$--spin model starting at a temperature $T=1.001\,\TMCT$ ($\TMCT=0.805166$) and quenching to zero. The new integration scheme allows us to reach much longer times than in previous works.}
\label{fig:1}
\end{figure}

The great advantage of the new integration scheme is evident from the results reported in Fig.~\ref{fig:1}. We can reach times of the order of $10^6$, that is, orders of magnitude larger than what can be achieved with the standard scheme of integration that uses a dense grid.
These large times are essential when extrapolating to the $t\to\infty$ limit.
For example, in the top panel of Fig.~\ref{fig:1} we plot $C(t,0)$ whose $t\to\infty$ limit is crucial in understanding which scenario is more likely, between the WEB and the SEB, the latter being the correct one as soon as $\lim_{t\to\infty}C(t,0)>0$.
This limit is easy to estimate by considering the large time results obtained with the new integration scheme, while it was harder to estimate with the shorter time results obtained in Ref.~\cite{folena2020rethinking}.

In the following, we are going to discuss the new physical behavior that arises from the integration at very long times of the DMFE for the spherical $(3+4)$--spin model with $f(q)=(q^3+q^4)/2$. We choose the starting configuration to be in equilibrium at temperature $T=1.001\, \TMCT$ and the following evolution to be a quench at $T_f=0$. This is a relevant physical situation, given that a realistic glass model can be thermalized above $\TMCT$. The further quenching mimics fast cooling in the glass phase. The results do not depend critically on the two temperatures (the initial one and the thermal bath one), so the behavior we are going to discuss is very generic. In a future publication, we will discuss other scenarios that arise by selecting different models and temperatures.

In the CK scenario, once the one-time quantities (e.g., the energy) reach the asymptotic value, the two-time observables keep aging, that is, they maintain a dependence on both times which is invariant under nonlinear time transformation \cite{kurchan2023time}. The key assumption that allows for the analytic solution of the DMFE is the decoupling between short and long times, which happens if WEB holds, that is, if $\lim_{t\to\infty}C(t,\tw)=0\;\forall \tw$. Forgetting the past is essential for the CK scenario to hold.

Fig.~\ref{fig:1} unequivocally shows that in the $(3+4)$--spin model we are studying, the WEB does not hold. Indeed, the correlation $C(t,0)$ does not go to zero, and the initial condition is not forgotten. The system keeps a memory of all its history and undergoes SEB.
This conclusion was already conjectured in Ref.~\cite{folena2020rethinking}, but the numerical evidence in that work was not solid enough and required extrapolations, given that the numerical integration was carried on with a dense grid up to times $O(10^3)$. We see in Fig.~\ref{fig:1} that both $C(t,0)$ and $R(t,0)$ level off to a plateau value only when times reach values $O(10^4)$.

The system's energy $E(t)$ exhibits faster dynamics as it relaxes towards its asymptotic value within timescales $O(10^2)$. We found excellent agreement when comparing the energy evolution with the data reported in Ref.~\cite{folena2020rethinking}. In that work, the best evidence against the CK scenario was provided by the energy relaxing below the threshold energy, given that only the energy achieved its asymptotic value on the time scales $O(10^3)$. In the present work, by reaching time scales $O(10^6)$, we achieve, for the first time, unquestionable evidence for SEB, by showing $\lim_{t\to\infty}C(t,0)>0$ \cite{folena2023weak}.

\begin{figure}[t]
\includegraphics[width=0.9\columnwidth]{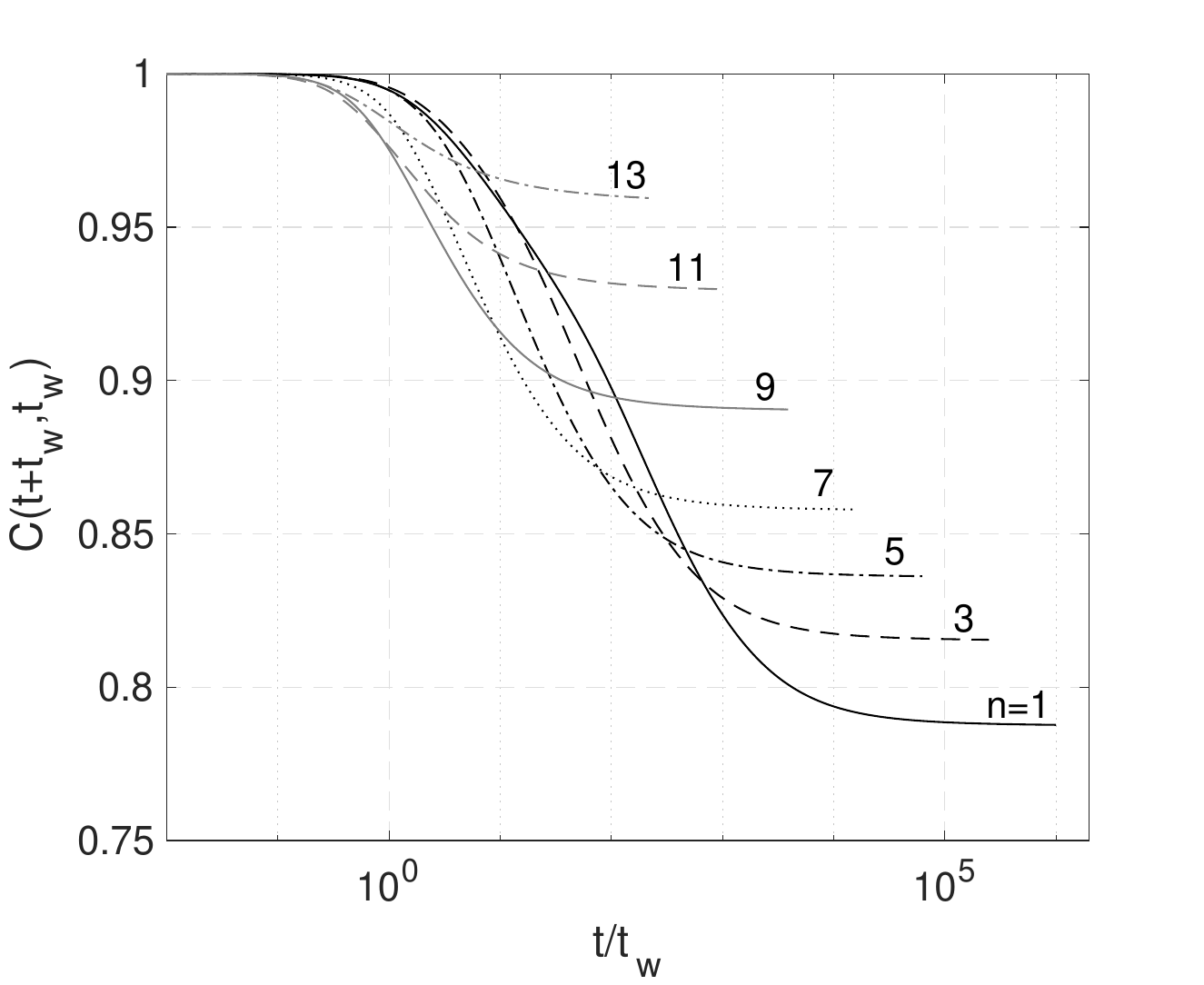}
\caption{The correlation $C(t+\tw,\tw)$ as a function of the rescaled time $t/\tw$ for a mixed $(3+4)$--spin model started at $T=1.001\,\TMCT$ and quenched to $T_f=0$. The curves are for different waiting times $\tw=2^n$. The system ages in a region of the configurational space, which shrinks as time passes.}
\label{fig:2}
\end{figure}

To better characterize the decorrelation process in the presence of SEB, we report in Fig.~\ref{fig:2} various relaxation profiles associated with different waiting times $\tw=2^n$ with $1\le n \le 13$. We plot $C(\tw+t,\tw)$ as a function of $t/\tw$ as is common in the study of the aging regime. We observe that a simple time rescaling is not enough to make curves collapse (as it happens in the pure spherical $p$--spin model \cite{kim2001dynamics}). Nonetheless, the system ages as it is evident that the relaxation time grows with $\tw$.
Two notable differences with respect to WEB aging are in order: (i) the correlation function always relaxes to a non-zero value for any $\tw$; (ii) the asymptotic value, $C(\infty,\tw)$, is an increasing function of the waiting time $\tw$ \cite{bernaschi2020strong}.

The new scenario emerging from our observations is that of a system that ages while exploring a bounded subregion of the configurational space \cite{capone2006off}. The size of this bounded region shrinks while time grows. This has profound consequences on the attempts to analytically solve this kind of out-of-equilibrium process: indeed, correlations with the past are important and must be introduced in the asymptotic ansatz.

\begin{figure}[t]
\includegraphics[width=\columnwidth]{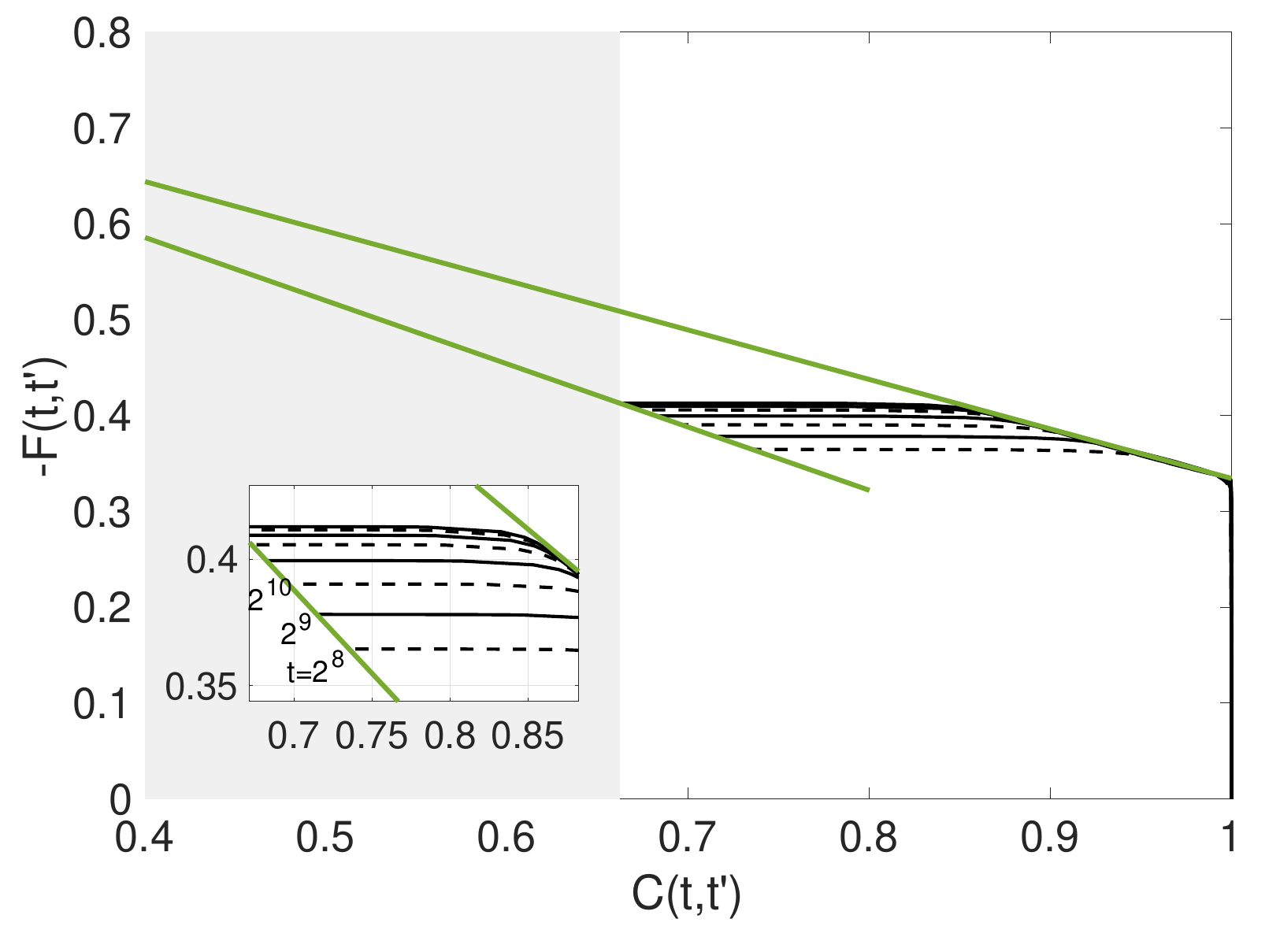}
\caption{Breakdown of the fluctuation-dissipation relation in a mixed $(3+4)$--spin model started at $T=1.001\,\TMCT$ and quenched to $T_f=0$. The gray region is inaccessible to the system dynamics as it is below the lowest achievable correlation $C(\infty,0)=0.658$. The curves are for different times $t=2^n$ and are plotted parametrically in $t'\in[0,t]$. Three regimes with different effective temperatures are visible in the plot. Data with the largest times are approaching the asymptotic fluctuation-dissipation curve (see inset).}
\label{fig:3}
\end{figure}

The aging process can be further investigated by looking at the connection between the correlation function $C(t,t')$ and the integrated response $F(t,t')$. We plot in Fig.~\ref{fig:3} the fluctuation-dissipation relation (FDR), a very well-known tool to inspect the out-of-equilibrium behavior in disordered models \cite{cugliandolo1994out,franz1995fluctuation,franz1998measuring,marinari1998violation,parisi1999generalized,ricci2003measuring}.
We remind the reader that for $T_f=0$, the self-overlap of every state is equal to 1, and thus the system is out-of-equilibrium for any $C(t,t')<1$. In other words, the equilibrium dynamics (satisfying the fluctuation-dissipation theorem) takes place only at $C=1$ and corresponds to the vertical part of the FDR at $C(t,t')=1$ in Fig.~\ref{fig:3}. For $C(t,t')<1$ the system is out-of-equilibrium.

In $p$--spin models undergoing aging with WEB, the FDR in the out-of-equilibrium regime looks like a straight line (whose slope is related to the effective temperature \cite{cugliandolo1997energy} and the Parisi parameter of states dominating the dynamics \cite{cugliandolo1993analytical}) down to $C=0$ \cite{kim2001dynamics,ricci2003measuring}.
In the mixed $(3+4)$--spin model we are studying, due to SEB, this can not happen as any correlation must satisfy the lower bound $C(t,t') \ge C(\infty,0) = 0.658$.
For this reason, we have shaded in gray the inaccessible region in Fig.~\ref{fig:3}.

Another important difference between WEB and SEB aging is the presence of a third dynamical regime in the FDR.
After the quasi-equilibrium regime (here concentrated in $C=1$) and the off-equilibrium regime with constant effective temperature (data on the upper green line), the system enters a third dynamical regime with zero effective temperature (the one with flat FDR) in Fig.~\ref{fig:3}.
This new regime was already observed in Ref.~\cite{folena2020rethinking}, but the timescales reached there were not enough to make any claim on the asymptotic behavior.

In the inset of Fig.~\ref{fig:3}, we can appreciate how curves for different $t$ values accumulate on the asymptotic large times FDR. Thanks to our new integration scheme, we can now safely assert the asymptotic FDR in mixed $p$--spin models is different from the one in pure $p$--spin models, with an effective temperature rapidly changing from the value already known from the pure model solution (the slope of the upper green line in Fig.~\ref{fig:3}) to a null value below a certain correlation threshold (that in the present case is $C_0 \simeq 0.87$). This new behavior must be incorporated in any new ansatz aiming at solving analytically the DMFE.

In summary, we have introduced a new integration scheme for the DMFE of spherical mixed $p$--spin models that allows us to reach time scales $O(10^6)$. At these very large times, the out-of-equilibrium dynamics of a system started in equilibrium slightly above $\TMCT$ and quenched at zero temperature can be fully understood, without relying on risky and noisy extrapolations (see Appendix \ref{sec:appA} for further evidence).
For the first time, we achieve unquestionable evidence for \emph{strong ergodicity breaking} in spherical $p$-spin models.
We find that the aging dynamics take place in a subregion of the configurational space that shrinks while time grows.
The fluctuation-dissipation relation gets modified with respect to the pure models, with the appearance of a new correlation range with zero (or very small) effective temperatures.

We believe our new numerical findings will stimulate the search for a more accurate ansatz to solve analytically the DMFE in the large time limit.
Moreover, thanks to the new integration scheme, it will be possible to study in detail the transition from the weak ergodicity breaking scenario to the strong ergodicity breaking one. The existence of such a phase transition was conjectured in Ref.~\cite{folena2020rethinking}, and is confirmed by the data in the present manuscript.
Very recently, a similar study \cite{lang2025numerical} that achieved comparable time scales using a different integration scheme, attempted to estimate the critical temperature $\TSEB$ of such a phase transition. While it is clear that, around their estimate, $\TSEB=1.0377(2)\,\TMCT$, the quantity $C(\infty,0)$ changes from being clearly larger than zero to being very close to zero, we believe it is not possible to reliably determine the critical temperature $\TSEB$ solely from the numerical data (especially with a such a small uncertainty). We provide evidence of this in Appendix \ref{sec:appB}. Probably an ansatz like the one used in Ref.~\cite{folena2020rethinking} (that provided an estimate $\TSEB \simeq 1.13\,\TMCT$) will be needed, and we leave this task for a future study.

\begin{acknowledgments}
This research has been supported by the ``National Centre for HPC, Big Data and Quantum Computing - HPC'', Project CN\_00000013, CUP B83C22002940006, NRP Mission 4 Component 2 Investment 1.4,  Funded by the European Union - NextGenerationEU.\\
\end{acknowledgments}

\appendix

\section{Three orders of magnitude more in time avoid risky extrapolations}
\label{sec:appA}

\begin{figure}[t]
\includegraphics[width=\columnwidth]{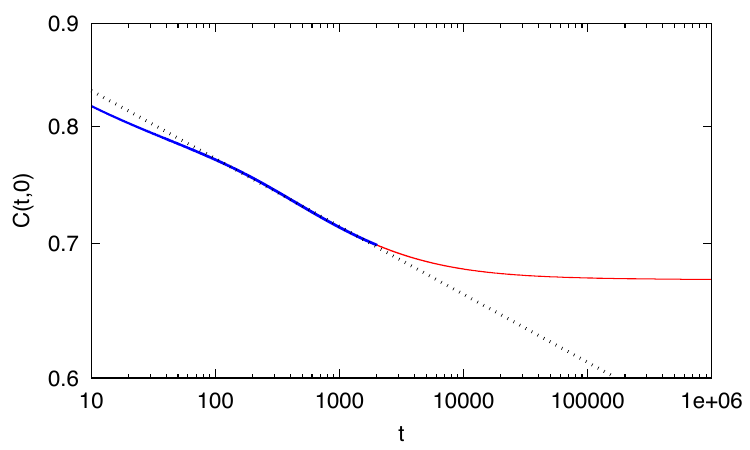}
\caption{Decay of the correlation $C(t,0)$ in a zero temperature relaxation, starting from equilibrium at $T=1.001\,\TMCT$ (same data shown in the upper panel of Fig.~\ref{fig:1}). The data marked with a blue thick line are those available prior to the present work, and a power law fit to the last decade of these data (dashed line) makes the WEB scenario hard to exclude. The new data collected in the present work extend to $t \sim O(10^6)$ and make the SEB scenario sure, without any extrapolation.}
\label{fig:appA}
\end{figure}

In Fig.~\ref{fig:appA}, we report the same data for $C(t,0)$ shown in Fig.~\ref{fig:1}, but here we highlight with a blue thick line the data that were available to the authors of Ref.~\cite{folena2020rethinking}. The data are shown on a doubly logarithmic scale to provide evidence that the last decade of the blue data could be reasonably interpolated by a single power law decay (the dashed line in Fig.~\ref{fig:appA}). Thus asserting, beyond any reasonable doubt, that the SEB scenario holds, based on data available prior to the present work, was somehow risky (indeed, the authors of Ref.~\cite{folena2020rethinking} focused on other observables and the direct claim $\lim_{t\to\infty}C(t,0)>0$ was not stressed much).
Instead, the new data obtained in the present work, extending to $t \sim O(10^6)$ make the SEB scenario certain, without any risky extrapolation in time.

\section{Difficulties in estimating $\TSEB$}
\label{sec:appB}

\begin{figure}[t]
\includegraphics[width=\columnwidth]{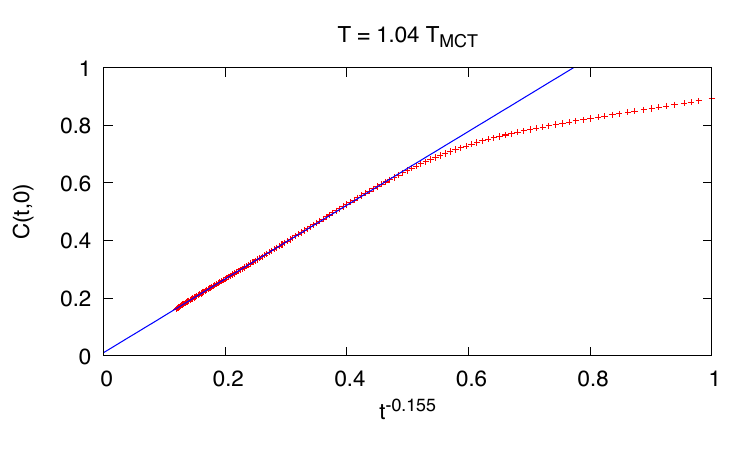}
\caption{The zero-temperature relaxation of $C(t,0)$ starting from $T=1.04\,\TMCT$, which is above the $\TSEB$ estimate provided in Ref.~\cite{lang2025numerical}, is compatible with an extrapolation to a non-zero value for $C(\infty,0)$. This observation strongly suggests that a very precise estimate of $\TSEB$ is out of reach with the present data.}
\label{fig:appB}
\end{figure}

The authors of Ref.~\cite{lang2025numerical} provide a very precise measurement of the critical temperature $\TSEB=1.0377(2)\,\TMCT$, that separates the high temperature region ($T>\TSEB$) where the WEB scenario holds and $C(\infty,0)=0$, from the low temperature region ($T<\TSEB$) where the SEB scenario holds and $C(\infty,0)>0$.

We believe that it is very unlikely that one can achieve such a precise estimate (with a relative uncertainty of $10^{-4}$) solely using the numerical data obtained from the integration of the DFME up to times of order $O(10^6)$.

To support our claim, we have integrated the DMFE corresponding to a zero-temperature relaxation that started from an equilibrium measure at temperature $T=1.04\,\TMCT$. This value is very close to, but definitely above, the estimate of $\TSEB$ obtained in Ref.~\cite{lang2025numerical}.
So, according to that work, we should find $\lim_{t\to\infty}C(t,0)=0$ with high statistical confidence (the temperature we are using is more than 10 standard deviations above their estimate for $\TSEB$).

In Fig.~\ref{fig:appB} we plot the data for the decay of $C(t,0)$ as a function of $t^{-0.155}$. The power law exponent is the one that better fits the data, under the assumption that the large time behavior is given by
\begin{equation}
\label{eq:fit}
    C(t,0) \simeq C(\infty,0) + A t^{-B}\;.
\end{equation}
Such an assumption seems valid for our data, given that the fit in Eq.~(\ref{eq:fit}) well interpolates $C(t,0)$ data in the range $t \ge 10^3$, with an exponent $B = 0.155$ and an extrapolated value $C(\infty,0)=0.0114(18)$. This value is small, but definitely non-zero.

While we do not want to make the claim that the above fit implies a SEB scenario for $T=1.04\,\TMCT$, since risky extrapolations are at work, we strongly believe that an estimate of $\TSEB$ solely from these numerical data is not accurate and should be postponed to future works.

\bibliography{biblio}

\end{document}